# Design of broadband high-efficiency superconducting-nanowire single photon detectors

L. Redaelli[1,2,*], G. Bulgarini[3], S. Dobrovolskiy[3], S. N. Dorenbos[3], V. Zwiller[1,2,4,5], E. Monroy[1,2], J.M. Gérard[1,2]

[1] Univ. Grenoble Alpes, F-3800 Grenoble, France
[2] CEA, INAC-PHELIQS, Nanophysics and Semiconductors group, F-38000, Grenoble, France
[3] Single Quantum B.V., 2628 CH Delft, The Netherlands
[4] TU Delft, Kavli Institute of Nanosciences, 2628 CH Delft, The Netherlands
[5] KTH Stockholm, Departement of Applied Physics, 114 28 Stockholm, Sweden
[*]luca.redaelli@cea.fr

**Abstract:** In this paper several designs to maximize the absorption efficiency of superconducting-nanowire single-photon detectors are investigated. Using a simple optical cavity consisting of a gold mirror and a $SiO_2$ layer, the absorption efficiency can be boosted to over 97%: this result is confirmed experimentally by the realization of an NbTiN-based detector having an overall system detection efficiency of 85% at 1.31 µm. Calculations show that by sandwiching the nanowire between two dielectric Bragg reflectors, unity absorption (> 99.9%) could be reached at the peak wavelength for optimized structures. To achieve broadband high efficiency, a different approach is considered: a waveguide-coupled detector. The calculations performed in this work show that, by correctly dimensioning the waveguide and the nanowire, polarization-insensitive detectors absorbing more than 95% of the injected photons over a wavelength range of several hundred nm can be designed. We propose a detector design making use of GaN/AlN waveguides, since these materials allow lattice-matched epitaxial deposition of Nb(Ti)N films and are transparent on a very wide wavelength range.

## 1. Introduction

Highly efficient detection of single photons is of capital importance not only for advanced quantum optics experiments but, increasingly, for many emerging industrial applications [1,2]. Single-photon detectors (SPDs) are used today in many fields, such as quantum key distribution [3,4], time-of-flight depth ranging applications such as the Lidar [5,6], space-to-ground optical communication [7], singlet oxygen detection for medical applications [8], and integrated circuit testing [9,10].

In recent years, superconducting-nanowire single-photon detectors (SNSPDs) have undergone rapid development, and they now outperform single-photon avalanche diodes (SPAD) [11,12], especially in the infrared spectral range. Very high efficiencies, up to 93% [13–15] have been demonstrated with SNSPDs at telecom wavelength, which, combined with the very high detector speeds, make SNSPDs the ideal candidate for many of the aforementioned applications [16].

The overall system detection efficiency ($\eta_{SDE}$) of a SNSPD can be seen as the product of three contributions: the coupling efficiency ($\eta_{coup}$), the absorption efficiency ($\eta_{abs}$) and the internal efficiency ($\eta_{int}$) [1]: $\eta_{SDE} = \eta_{coup} \times \eta_{abs} \times \eta_{int}$.

The coupling efficiency $\eta_{coup}$ is the probability that a photon travelling in the optical fiber carrying the signal is coupled to the detector. The absorption efficiency $\eta_{abs}$ provides a measure of the probability, for a photon impinging on the detector, to be actually absorbed by it rather than transmitted or reflected. Finally, the internal efficiency $\eta_{int}$ is the probability of registering a voltage pulse at the nanowire leads when a photon is absorbed. This latter parameter depends on the choice and quality of the superconducting material (reported values reach > 90% for Nb(Ti)N [17], and close to 100% for WSi [13]), and its optimization will not be discussed in this paper.

The coupling efficiency, $\eta_{coup}$, is usually boosted by folding the detecting nanowire in a tightly packed meandering structure, covering a circular area several microns in diameter. It is then possible to efficiently couple the light output from a fiber to the center of the nanowire meander with a self-aligning method, achieving coupling efficiencies close to unity, [18].

Achieving high absorption efficiency is often difficult in SNSPDs. Since the light is usually coupled vertically to the thin (4-8 nm) nanowire, only a small fraction of it is absorbed. Several techniques have been developed to enhance $\eta_{abs}$. In particular, two device designs have been proposed: the "microcavity-enhanced" design, in which the meandering wire is enclosed in an optical cavity to enhance absorption [Fig. 1(a)] [13–15,19–23], and the "waveguide-coupled" (or "travelling-wave") design, in which the nanowire is patterned on top of a waveguide [17,24–27]. The light is injected and propagates in the waveguide, and the evanescent tail of the mode couples to the nanowire, eventually leading to complete absorption [Fig. 1(b)].

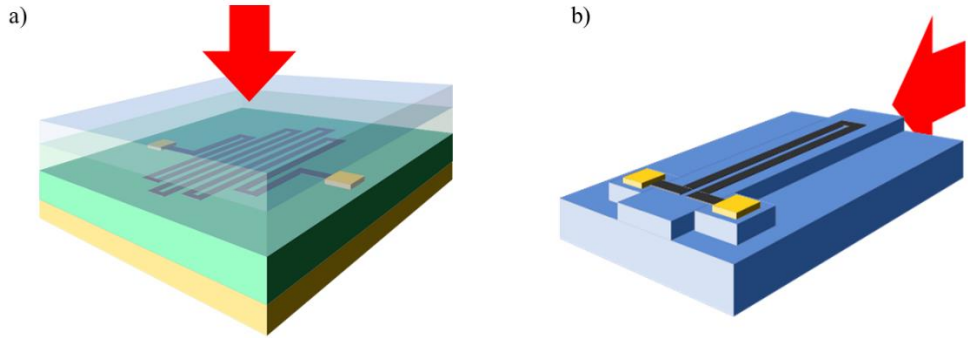

Fig.1 Illustration of the device designs investigated in this paper: in (a) the microcavity-enhanced detector, in (b) the waveguide-coupled detector (The drawings are not to scale).

In this paper, both approaches will be studied by optical simulations, and their respective advantages and drawbacks will be discussed. First, the microcavity approach will be studied (section 2), with several designs of increasing complexity. One of the designs has been realized experimentally and characterized, in order to validate the calculations. More complex cavity designs enabling absorption efficiencies close to unity will be then presented. Finally, high-efficiency polarization-insensitive detectors based on the waveguide-coupled design will be studied in section 3. An innovative device architecture will be proposed, in which the nanowire is deposited on an AlN/GaN waveguide.

## 2. Microcavity-enhanced detectors

*2.1 Simulation method and optical model*

In this section, devices based on the microcavity approach are studied by 2D finite-difference time-domain (FDTD) simulations using the commercial software *RSoft FullWAVE* [28]. The device is modeled as an infinite grating, as shown in Fig. 2(a). Only one grating period is drawn and calculated, and in-plane periodic boundary conditions are applied. The 2D analysis is justified because the photons are incoming at normal incidence and the wavefront can be approximated as planar: the grating period is much smaller than both the wavelength and the detector diameter. The top and bottom domain boundaries are 0.2 µm-thick perfectly matched layers (PML). A nonuniform graded grid is used, in order to allow precise modeling of the thin nanowire but limit the overall calculation effort. The finest grid size, used at each layer boundary, is set to 1 nm in *x* direction, and to one tenth of the nanowire thickness (i.e. 0.7 nm) in *z* direction. The coarser grid size is 10 nm, both in *x* and *z*. For simplicity, throughout this section we will refer to "TE-polarized light" when the electric field is parallel to the nanowire length, "TM-polarized" light for the opposite case, where the electric field is perpendicular to the nanowire.

The absorption efficiency is calculated, once the steady state is reached, by subtracting from the input power the power reaching the top PML boundary due to reflection, and the power absorbed by the backside mirror (if any). The thickness of the backside mirror is chosen so that the transmitted power reaching the bottom boundary is < 0.1% of the input power.

The complex refractive indices used in the calculations were $4.80 + i6.05$ for NbTiN [29,30], $4.58 + i3.66$ for WSi [13], 2.45 for $TiO_2$ [31–33], 1.44 [34] for $SiO_2$, and $0.57 + i9.66$ for Au [34].

*2.2 Results and discussion*

The nanowire thickness is fixed to 7 nm, the width to 100 nm, and the spacing between two nearby nanowire meanders is set to 100 nm: values similar to these are typical for Nb(Ti)N based detectors [15, 35, 36]. In this configuration, assuming a semi-infinite SiO$_2$ layer, the calculated absorption probability would be as low as 29% for TE-polarized photons, and 11% for TM-polarized photons.

The absorption efficiency can be increased by placing a highly reflective mirror below the SiO$_2$ layer and reducing the SiO$_2$ thickness close to a quarter of the wavelength, so that light coupling in the nanowire is enhanced by constructive interference. In the real device, which makes use of a gold mirror and is designed for operation at telecom wavelength (1.55 µm), the phase changes induced upon reflection at the metallic/dielectric interfaces, as well as the non-zero NbTiN thickness, place the optimum SiO$_2$ thickness value for maximum absorption at 268 nm. Figure 2(b) displays the absorption in the superconducting nanowire as a function of the incident wavelength considering TE- and TM-polarized light. In this case, 97.6% absorption efficiency for TE-polarized photons is achieved, whereas TM absorption is much less efficient, with a peak value of 37%. Note that the maximum achievable efficiency would be lower, for both TE and TM, if a superconducting material with lower absorption coefficient (e.g. WSi) or a thinner nanowire are used.

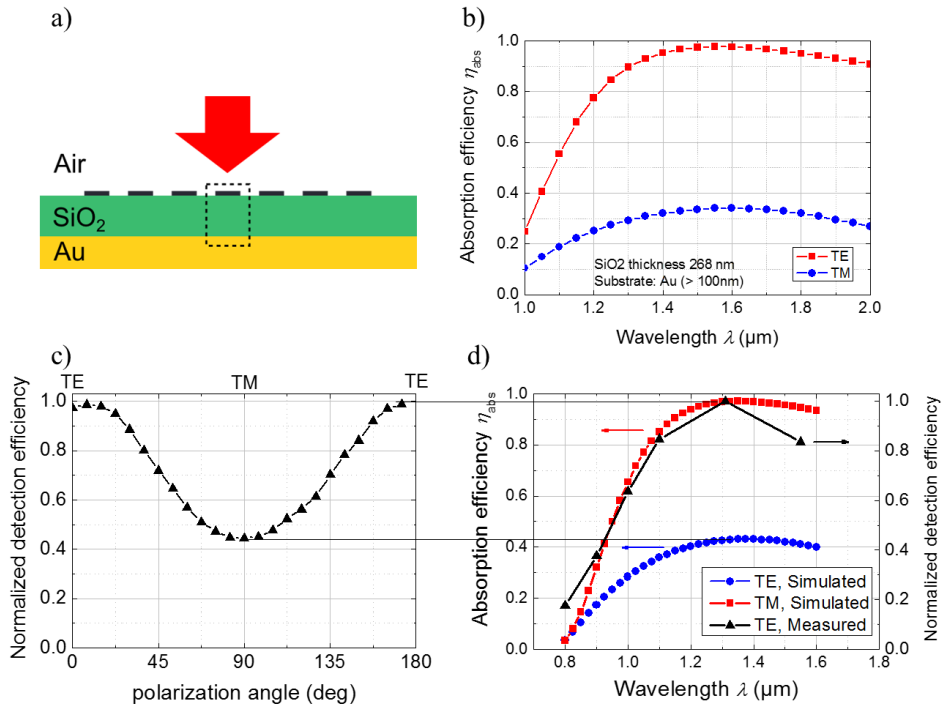

Fig. 2(a) Cross-section scheme of a "half-cavity" detector design (not to scale). The area enclosed in the dashed rectangle corresponds to the simulated unit cell. (b) Calculated absorption efficiency as a function of wavelength for the "half-cavity" device, where the SiO$_2$ thickness is 268 nm and the substrate is a 100 nm-thick Au mirror. In (c) and (d), measurements performed on devices optimized for peak absorption at 1.31 µm are shown. The normalized system detection efficiency is plotted as a function of photon polarization (c) and wavelength (d), and compared to simulation. The system detection efficiency before normalization in (d) reaches 85% at 1.31 um.

To validate these calculations, a device consisting of a NbTiN nanowire on SiO$_2$ and Au was fabricated and characterized by the company *Single Quantum*. The detection efficiency was measured at a temperature of 2.5 K. As photon source a *Fianium Supercontinuum Laser SC-400-4* with acousto-optic tunable filters was used; the optical pulses were attenuated by a *JDS Uniphase JDSU HA9* calibrated programmable attenuator, and the photon polarization was controlled by a *Thorlabs FPC561 Fiber Polarization Controller*. Electrical amplification of the detection pulses and counting were performed with a *Single Quantum* proprietary driver, while optical power measurements were performed by a *Thorlabs Digital Optical Power and Energy Meter PM100D* equipped with a Ge photodiode. The measurement values at 1.31 µm, wavelength at which the structure was optimized to operate, were 10 nW average laser power with an additional attenuation of 50 dB, which correspond to a photon flux of 6.59 × 10$^5$ photons/second.

The measurements results are shown in Fig. 2(c) and 2(d), together with the simulated curves. The system detection efficiency $\eta_{SDE}$, reached 85% at 1.31 µm, which is, to the best of the authors' knowledge, the highest reported to date for Nb(Ti)N-based SNSPDs. We attribute the difference between this value and the calculated estimate of the maximum absorption efficiency ($\eta_{abs}$= 97.0%) mainly to the limited internal efficiency $\eta_{int}$ of these devices. In order to compare experimental measurements and simulations, we assume that the coupling efficiency and the internal efficiency are constant across the considered wavelength range, and independent from light polarization. We can thus normalize the measured system detection efficiency to the maximum absorption efficiency and directly compare the two curves. The good agreement between measurement and simulation confirms that our model predicts accurately the spectral and polarization dependence of the SNSPDs under investigation.

It is possible to further increase the absorption efficiency of the devices by adding a matching dielectric mirror on the device front side, thereby completing the optical cavity. By optimizing the thickness of the dielectric layers, the reflected power can be reduced close to zero thanks to destructive interference, so that all the power is absorbed inside the cavity. (i.e., $\eta_{abs} \sim 1$). If the backside reflector is a gold mirror and the previously introduced nanowire meander sizes are used, the maximum efficiency that can be achieved is 98.6%. In order to achieve unity efficiency (> 99.9%) the Au mirror must be replaced by a dielectric Bragg mirror, since Au is not a perfect reflector at 1.55 μm—it absorbs at least 1–2% of the incident radiation. In Fig. 3 both cavity configurations are illustrated, and the respective calculated absorption efficiencies are plotted as a function of wavelength and polarization. In comparison to the "half-cavity" configuration in Fig. 2, TM absorption is enhanced, but the absorption peak is blue shifted with respect to the TE absorption peak. By tuning the front mirror and spacer thicknesses, it is possible to design a polarization-independent detector, although this will come at the cost of peak efficiency. For example, calculations show that for front mirror thicknesses of 263 nm ($TiO_2$) and 89 nm ($SiO_2$), the absorption efficiency would be of 80% for both polarizations.

In these "full-cavity" configurations, peak efficiencies close to unity can be reached at any targeted wavelength and photon polarization, almost independent of the thickness of the $SiO_2$ layer below the superconducting nanowire and of the nanowire size and absorption coefficient, by tuning the reflectivity of the front mirror. This fact opens up novel interesting opportunities from a fabrication point of view: devices from the same wafer could be tuned to maximize the response at different wavelengths (or photon polarizations) at the end of the fabrication process, by fabricating front mirrors having different $SiO_2$ and $TiO_2$ thicknesses.

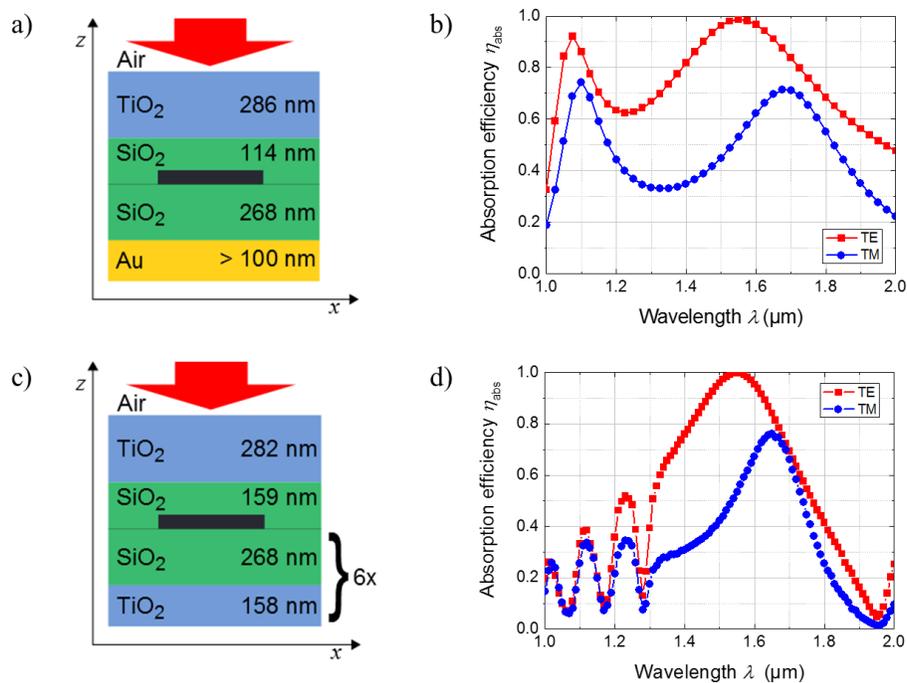

Fig. 3 Illustration (not to scale) of the simulated unit cells of microcavity detectors optimized for high TE absorption efficiency at 1.55 μm [(a): backside Au mirror, (b): backside Bragg mirror)]. The respective efficiencies, calculated as a function of wavelength and polarization, are shown in (b) and (d).

Microcavity-enhanced detectors are intrinsically wavelength sensitive, since they make use of interference effects to enhance absorption. When using a simple backside mirror ("half-cavity", Fig. 2), the detection band is quite broad, with a 0.7 µm-wide wavelength range around the peak where the absorption efficiency exceeds 90%. When incorporating a front mirror [Fig. 3(a)], on the other hand, the detection band (absorption > 90%) drops to below 0.2 µm, and it shrinks even further when using a backside Bragg mirror.

In order to achieve polarization-insensitive and broadband detectors, a change of paradigm is necessary. A possible approach is the waveguide–coupled detector design, which will be discussed in the next section.

## 3. Waveguide-coupled detectors

*3.1 Detector design options*

Various substrate/waveguide material combinations have been proposed for the realization of waveguide-coupled SNSPDs. The most obvious choice is silicon-on-insulator (SOI), in order to integrate this detector technology on-chip with photonic circuits working at telecom wavelength. Silicon waveguides rely on a very mature technology, with extremely low propagation losses [37]. Furthermore, the index difference between silicon and the underlying $SiO_2$ cladding is very high (about 2), and the mode can be tightly confined in very thin waveguides, providing intense evanescent field. This, in turn, allows for efficient evanescent coupling to the nanowire and highly efficient absorption, even in short waveguides: attenuations up to 1 dB/µm have been demonstrated [27]. The main drawback of this technology is that silicon is not transparent below 1.1 µm, so that it is not suitable e.g. for 850 nm fiber-optics transmission or for truly broadband detectors.

Other approaches include GaAs/AlGaAs [25] and $Si_3N_4$/$SiO_2$ [24] waveguides. In both cases the index difference is much lower than for SOI, which means that the mode confinement is less effective. Therefore, thicker and wider waveguides are needed to guide the mode, and its evanescent field will be less intense, so that longer detectors are necessary to achieve absorption efficiencies close to unity. The GaAs absorption edge is located around 0.9 µm, which allows only slightly larger bandwidth than with silicon, while high-quality $Si_3N_4$ can have a much wider transparent band.

In this work we propose waveguide-coupled detectors fabricated on GaN/AlN waveguides, since this material system has some important advantages: AlN is lattice-matched to Nb(Ti)N, offering the best possible substrate for the deposition of high-quality superconducting films [38]. Furthermore, AlN and GaN are transparent over a very large band, from 0.4 µm to about 7.4 µm [39]: this opens the way to the design of detectors which can operate with high efficiency on a very broad wavelength range. The ridge waveguide design is illustrated in Fig. 5(a). It consists of a double GaN/AlN waveguiding layer on a sapphire substrate. In this configuration, the mode maximum can be efficiently confined in the upper GaN layer. Sapphire acts as a cladding layer, thanks to the larger index difference with AlN (at 1.55 µm, sapphire 1.75 [31,40], AlN 2.12 [31,41]) and GaN (2.32 [31,42]). The thick AlN layer is needed as a buffer in order to achieve high crystalline quality in the upper GaN layer, where the maximum mode intensity propagates, and has a thickness of 1.1 µm. Besides, the overall thicker waveguiding layers should facilitate fiber-waveguide coupling. A cleaved waveguide facet is shown in Fig. 4(b).

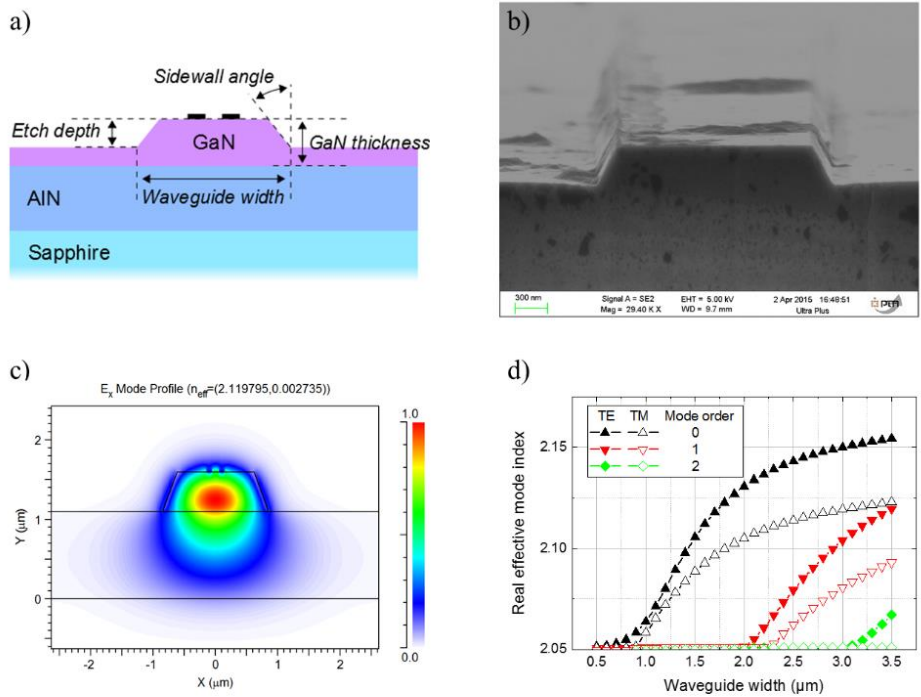

Fig. 4(a) Illustration of the waveguiding structure in cross-section (not to scale). The labels in italics indicate the most important parameters that will be optimized in the following. Note that the sidewall angle is defined so that a perfectly vertical sidewall would have a 0° angle. (b) SEM micrograph of a GaN/AlN ridge waveguide. The etch depth is here 0.35 µm, the sidewall angle about 25°. The smooth facet is obtained by cleaving along the natural GaN cleaving plane (m-plane). (c) Calculated fundamental TE mode for GaN thickness = etch depth = 0.5 µm, sidewall angle = 25°. d) Effective index of the first three TE and TM modes as a function of waveguide width (same waveguide geometry as before). The waveguide supports only the fundamental TE and TM mode for widths between 1 µm and 2 µm.

*3.2 Simulation method and optical model*

In order to quantify the absorption efficiency, we calculate the mode profile and complex effective index using a Finite Element (FEM) mode solver (RSoft FemSIM [28]), as shown in Fig. 4(c). The nanowire absorption is calculated from the imaginary part of the complex effective mode index. This method does not take into account the mode reflection at the nanowire onset, nor the additional absorption by the short transversal wire segment connecting the two parallel wire segments. This approximation is justified by the fact that the contribution of both effects is negligible, as pointed out in Ref. [27] and confirmed by a 3D FDTD simulation performed by the authors, in which these contributions were smaller than the numerical noise level of the simulation.

The nonuniform graded mesh is chosen so that the finest mesh size is ten times smaller than the smallest structure, both in x and z directions (0.7 nm in $x$, 10 nm in $z$). A coarser mesh is used (0.02 µm) where the material is uniform, far away from the material interfaces. The nanowire dimensions and spacing are unchanged from the microcavity design in section 2 (7 nm thick and 100 nm wide nanowire, with 100 nm spacing).

The goal of the simulations is to optimize the waveguide dimensions in order to ensure single-mode operation and maximize the modal absorption. The most important parameters are highlighted in Fig. 4(a): the waveguide width and thickness, the ridge waveguide etch depth, and the sidewall angle. Finally, the minimum waveguide length necessary to achieve unity absorption will be determined, considering the impact of the mode propagation losses.

*3.2 Results and discussion*

First, we define the maximum waveguide width which only supports the fundamental TE (electric field parallel to the surface) and TM (electric field perpendicular to the surface) modes. The calculation results, displayed in Fig. 4(d), show that the first-order TE mode starts oscillating for waveguide widths larger than 2 µm: this will be the upper limit for the waveguide width.

The various waveguide dimensions (ridge width, GaN thickness, etch depth and sidewall angle) are then investigated in order to maximize absorption. In Fig. 5(a) and (b) the modal absorption of the fundamental TE and TM modes is plotted as a function of the waveguide width and the GaN thickness, respectively. Since TE absorption is consistently lower than TM absorption, the primary goal is to maximize TE absorption. The optimum is reached for a ridge width of 1.7 µm and a GaN thickness of 0.5 µm. If the waveguide size is smaller, the maximum mode intensity is pushed down in the AlN layer, and the intensity of the evanescent mode tail drops, i.e. the modal absorption drops. If, on the contrary, the waveguide is too large, the mode is more tightly confined in the GaN layer, thus reducing the evanescent tail intensity.

In Fig. 5(c) and (d) the ridge etch depth and its sidewall angle are investigated. Deeper waveguides offer better lateral confinement, thus more intense evanescent field, up to an etch depth of 0.8 µm. The sidewall slope has almost no impact on the modal absorption, up to an angle of 35° [Fig. 5(d)]. On the other hand, a deeper etch would also increase the propagation losses, due to the increased interaction of the mode with the rough etched sidewalls. The sidewall angle is indirectly linked with the propagation losses, as well: the choice of the ICP etching parameters influences both the surface roughness and the sidewalls slope. Therefore, before choosing etch depth and angle, the impact of the waveguide losses on the device efficiency should be estimated. In this work we arbitrarily decide to stop the etching at the GaN/AlN interface (etch depth = 0.5 µm), where TE absorption starts to saturate, and consider a 25° sidewall angle, as obtained experimentally in Fig 4b.

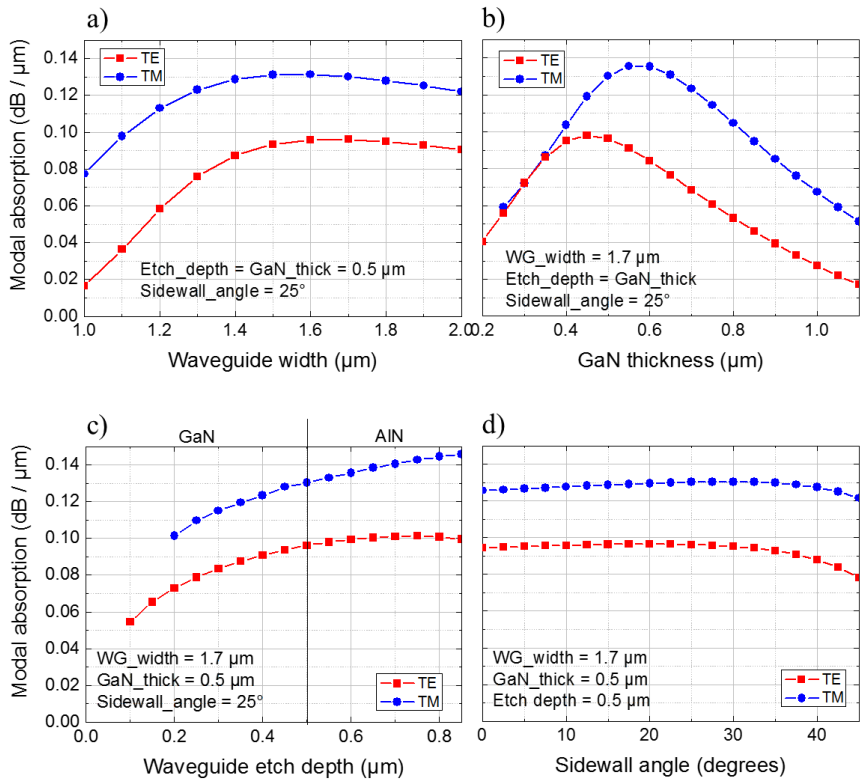

Fig. 5: The modal absorption coefficient (in dB/µm) is plotted as a function of the waveguide width (a), GaN thickness (b), ridge etch depth (c), and waveguide sidewall angle (d). Each time, only one parameter is varied; the three other parameters are fixed and their values are indicated in each plot.

For a bare waveguide structure as the one considered here, propagation losses of about 2 dB/mm (4 dB/mm) or lower have been reported for TE (TM) polarization [43,44]. These values correspond to 2.1% (3.1%) of the TE (TM) peak modal absorption induced by the NbTiN nanowire, which is 0.096 dB/µm (0.130 dB/µm). As a consequence, the maximum TE (TM) absorption efficiency will be limited at 97.9% (96.9%) for an infinite waveguide length. In Fig. 6(a) the attenuation of the TE and TM fundamental modes, propagating in the waveguide, is plotted as a function of the waveguide length $L$. The power drops to 0.1% of the input power (-30 dB attenuation) for $L$ = 306 µm, i.e. an absorption efficiency of about 97.8% is achieved. TM attenuation, for the same $L$, is -10 dB larger, thus the maximum TM absorption efficiency of 96.9% is (almost) reached.

In Fig. 6(b), the absorption efficiency of the fundamental TE and TM modes is plotted as a function of wavelength. The curves are calculated for $L$ = 306 µm, assuming that the waveguide losses are equal to 2 dB/mm for TE and 4 dB/mm for TM across the whole wavelength range. It should be noted that, for wavelengths shorter than 1.2 µm (1.3 µm), the first-order TE (TM) mode is supported. As a consequence, with decreasing wavelength an increasingly large fraction of the radiation power might be injected in the first order mode, which is poorly coupling to the nanowire, i.e. the absorption efficiency below 1.3 µm is overestimated. Nevertheless, comparing Fig. 6(b) to Fig. 3(c) and (d), it can be stated that the waveguide approach offers high efficiencies on a much larger band, when compared to microcavity-enhanced detectors, and polarization sensitivity is negligibly small.

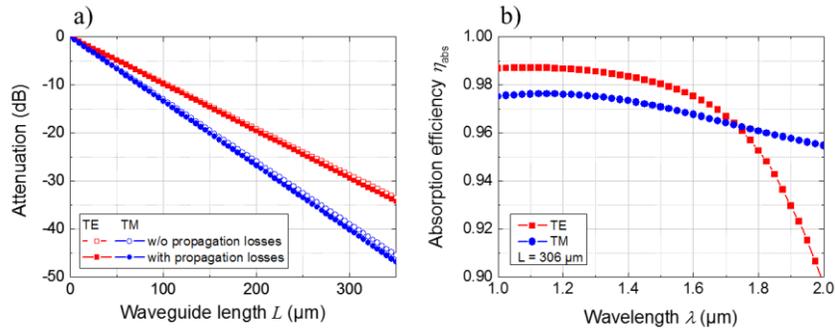

Fig. 6 (a) Mode attenuation as a function of the waveguide length. (b) Absorption efficiency (net of waveguide losses) as a function of wavelength. Constant losses of 2 dB/mm (4 dB/mm) for TE (TM) radiation are considered. Note that the y axis starts at 0.9, and not 0 as in Fig. 3.

A major drawback of the waveguide-coupled device design is that, in order to achieve high coupling efficiency, extremely precise alignment is needed. Here, the simple fiber self-aligning method developed for normal-incidence detectors described in Ref. [18] cannot be used. This is not an issue when thinking of waveguide-coupled detectors for integration in photonic circuits, where both the photon source and the detector are on the same chip: on-chip photonics is indeed a major application field for this kind of detectors, and has been motivating their development in the first place [17,24–27,36]. On the other hand, if the goal is the fabrication of a stand-alone detector for other industrial applications, the overall detector efficiency will be necessarily limited by the fiber-waveguide coupling efficiency. Coupling efficiencies below 1 dB can be achieved, and have been realized on Si CMOS technology [45], but they require complex and expensive alignment systems. The task is made even more difficult for SNSPDs, since they operate at cryogenic temperatures, and the induced mechanical strain due to thermal mismatch during cooling needs to be managed.

Finally, for applications where large coupling losses (i.e. low system detection efficiencies) are acceptable, but detector speed, polarization insensitivity, and/or broadband operation are essential, the waveguide approach might be interesting. Since the absorption efficiency depends logarithmically on the nanowire length, devices with short nanowires, i.e. shorter dead times and higher detection rates, can be designed. An interesting "hybrid" approach, combining waveguide-coupling and cavity effects has been recently proposed [36], in which high absorption efficiencies are achieved using ultrashort nanowires.

## 4. Conclusion

In this paper, different ways to enhance the absorption efficiency of SNSPDs have been investigated by optical simulations. Two main approaches have been discussed: the microcavity-enhanced device, where light absorption is boosted by embedding the superconducting nanowire meander in an optical cavity, and the waveguide-coupled device, where the evanescent tail of the mode is coupled to a nanowire deposited on top of the waveguiding structure.

The simplest microcavity design makes use of a backside gold mirror and a $SiO_2$ dielectric layer to enhance absorption in the 7 nm-thick NbTiN nanowire through constructive interference. High absorption efficiencies of 97.6% can be achieved at 1.55 µm, but only for linearly polarized light where the electric field oscillates parallel to the nanowire segments. A detector based on this architecture has been realized and characterized, and the measurements are consistent with the simulation results (peak system detection efficiency of 85% at 1.31 µm).

By adding a front dielectric mirror to complete the vertical cavity, an absorption efficiency of 98.6 % can be achieved at 1.55 µm. This value can be further increased to over 99.9% by replacing the backside Au mirror with a dielectric Bragg mirror. In both cases, the TE/TM polarization absorption ratio is between 1.4 and 1.7 but, by tuning the front mirror layer thicknesses, it is possible to reach a compromise and achieve detectors which are polarization-insensitive at 1.55 µm and have around 80% absorption efficiency.

Concerning the waveguide-coupled approach, realization on a GaN/AlN waveguide has been proposed as a good candidate for on-chip photonics applications. In this configuration, highly efficient and almost polarization-insensitive photon absorption over a very broad wavelength range is possible. The detector is designed to absorb over 97% of both TE- and TM-polarized photons at 1.55 µm, and over 90% of the photons over a wavelength range of more than 700 nm.

## Acknowledgments

This work was funded by The European Commission via the *Marie Skłodowska Curie IF grant* "SuSiPOD" (H2020-MSCA-IF-2015, #657497), the *French National Research Agency* via the "WASI" (ANR-14-CE26-0007) and "GANEX" (ANR-11-LABX-0014) programs, and the Grenoble *Nanosciences Foundation*.